\documentclass[doublecol]{epl2}

\title{The asymmetric Kerr metric as a source of CP violation}
\author{Mark J Hadley }

\institute{
  Department of Physics, University of Warwick, COVENTRY, CV4 7AL, UK
  \thanks{E-mail: \email{mark.hadley@warwick.ac.uk}}
}

\pacs{11.30.Er}{Charge conjugation, parity, time reversal, and other discrete symmetries}

\abstract{All experimental evidence for violation of discrete spacetime symmetries: Parity and Time reversal and the related Charge conjugation/Parity combination (P, T and CP respectively) has been obtained on earth in a gravitational potential that is P and T anisotropic. It is suggested that the origin of the observed CP violation is the scalar field equal to the frame dragging term $d\phi dt$ in the Kerr metric of a spinning massive body. The galaxy would be the largest such source. Indirect evidence of such an effect would be anisotropic decay products when plotted in a reference frame defined by the fixed stars. As a consequence, CP violation would be very much greater near compact astrophysical objects with large angular momentum.}

\begin{document}

\maketitle

\section{Introduction}
CP violation has been observed in the neutral Kaons \cite{wu} and $B^0$ and $D^0$ meson decays\cite{BABAR,Belle}. The results are accounted for in the standard model by a complex phase in the Cabibbo – Kobayashi – Maskawa, CKM, matrix (see for example \cite{harrison:1998}) which measures the mixing between the quark generations. The standard model appears to correctly parameterise CP violation but it does not account for the origin of the effect. In any other branch of physics we would not be satisfied without an external explanation for any asymmetry. Indeed, when CP violation was first discovered in the neutral Kaons an external galactic vector field was considered, but was discounted because a vector field would give rise to an energy dependent CP violation contrary to experimental results \cite{PhysRevLett.13.348, sudarsky}.

According to the theory of general relativity a rotating mass creates an asymmetric gravitational potential which can be described in general relativity by the Kerr metric (see for example \cite{MTW}). In more colourful language it drags spacetime around as it spins. This paper explores the possibility that the asymmetric metric is the source of the observed CP violation and suggests how existing data can be analysed to test for such effects.

Motivation exists at a several levels. Most simply, there is an asymmetry in the experimental arrangements which could and should be tested for. However deeper motivation comes from particle physics phenomenology and some routes for unifying general relativity and quantum theory.

Conventionally, particles are \emph{assumed} to be symmetrical under a parity operation (the particle wavefunctions are eigenstates of the parity operator). With that convention the weak interactions show maximal parity violation. The alternative is known but rarely discussed - namely that the weak interactions conserve parity and that particles themselves are intrinsically anti-symmetric, with the mirror image of a particle being its anti-particle \cite{griffiths}. In other words, that CP rather than P is the appropriate operator corresponding to an inversion of the space coordinates. With the alternative convention, particles are intrinsically asymmetric and we might reasonably expect particles and anti-particles to behave differently in an asymmetric potential.

General relativity is intrinsically parity conserving. This can be seen from the basic equation $ G_{\mu\nu} = 8 \pi G/c^4 T_{\mu\nu}$ which is a tensor equation equating two tensors of the same form. Parity (or time reversal) violation would require a tensor to be equated with a pseudo tensor, so that left and right hand side of the equation transform differently under reflections - the mirror image equation would then be different and only one would describe the world we live in. Although that might seem strange it is proposed by some researchers as a way to introduce parity violation into the strong interactions by adding a pseudoscalar term to a scalar Lagrangian (see for example \cite{Abel2001151}). The symmetric nature of general relativity should also be evident because it is a geometric theory formulated using differential geometry. Curved spacetime does not admit an intrinsic co-ordinate system and in general it is not possible to cover an arbitrary topology with only one well behaved co-ordinate system - instead we use a family of overlapping charts to parameterise the space. Applying the techniques of differential geometry derives results and equations that are independent of any particular co-ordinate system and which certainly do not distinguish between left or right handed charts. Of course chiral solutions may well exist, but for every such solution the mirror image would be an equally valid solution. It is also possible to have asymmetric solutions by imposing asymmetric boundary conditions.

A deeper motivation comes from attempts to combine a geometric theory like general relativity with particle physics. Because general relativity is intrinsically parity conserving, attempts at unification based on general relativity suggest that the observed parity violation is not a fundamental property of nature. This supports the unconventional parity assignments to particles, but also requires an explanation for the smaller CP violation effects. The external gravitational potential could in principle provide just such a mechanism, by providing asymmetric boundary conditions at a local level. Attempts to link the quantum theory and the internal structure of elementary particles with the classical background spacetime geometry, such as \cite{hadley2000}, will naturally have effects where asymmetry in the background metric affects particle properties.

\section{The gravitational potential of a rotating mass}

The CP violation experiments all take place in an asymmetric gravitational potential generated by the rotation of the Earth, Sun and Galaxy. In general relativity, gravitation is manifested as a curvature of the space and time. Given any coordinate system the curvature depends upon the metric tensor which is used to calculate lengths of space and time intervals. The metric tensor, $g_{\mu \nu}$ (where $\mu$ and $\nu$ vary over time and three space indices) is the generalisation of a scalar product to four dimensions, and curved geometry, in the given coordinates.

In the usual spherical coordinates, a spherical mass, $M$, gives rise to the Schwarzschild metric
\begin{eqnarray*}
ds^2 &=& -(1 - r_s/r) c^2 dt^2 + (1- r_s/r)^{-1} dr^2 \\
&& + r^2 \left(d\theta^2 + \sin^2\theta \, d\phi^2\right)
\end{eqnarray*}
where $r_s = 2 G M/ c^2$ is the Schwarzschild radius \cite{MTW}. Of particular note is the first term, which gives a gravitational redshift because the coordinate time interval $dt$ is greater than the proper time interval, $d\tau$,  measured by a local observer for whom $ds^2 = -c^2 d\tau^2$. It can be seen that the Schwarzschild metric is spherically symmetric and also does not depend on the time orientation because only the diagonal terms, ($g_{tt}, g_{rr} , g_{\theta\theta} , g_{\phi\phi} $), are non-zero and there is no explicit $t$ dependency. The coordinates used in the Schwarzschild solution $(t,r,\theta,\phi)$ are chosen because they show the symmetry of the geometry as clearly as possible and match up with the usual spherical spacetime coordinates at large $r$ where spacetime is asymptotically flat.

A rotating body has an axis of symmetry, rather than spherical symmetry, and the angular momentum of the source drags spacetime around with it. Using coordinates designed to show the symmetry, the metric for a given source mass $M$ with angular momentum $J$ takes the Kerr form
\begin{eqnarray*}
ds^2 &=& -(c^2-2 G M r/\rho^2)dt^2 +\rho^2/\Delta^2 dr^2+ \rho^2 d\theta^2 \\
  &&+ (r^2 + \frac{J^2}{M^2 c^2} + \frac{2 G J^2r}{c^4\rho^2 M}\sin^2\theta)\sin^2\theta d\phi^2 \\
&& + \frac{4G r J }{c^2 \rho^2 }\sin^2\theta dt d\phi  ,
\end{eqnarray*}
using the abbreviations
\begin{equation}
\rho^2 = r^2 + \frac{J^2}{M^2c^2} \cos^2\theta
\end{equation}
and
\begin{equation}
\Delta^2 = r^2 - \frac{2 G M }{c^2}r + \frac{J^2}{M^2 c^2} .
\end{equation}
The Kerr metric is equal to the Schwarzschild metric for zero angular momentum and has many subtle features for rotating black holes. For less extreme objects, the terms of the Schwarzschild metric are present but slightly modified; however, the Kerr metric adds a P and T violating term $g_{t\phi}dt d\phi$. Taking the weak field, non-relativistic, limit: $r \gg r_s$, and $J \ll Mr c$ gives the P and T asymmetric term:
\begin{equation}
g_{t\phi} \simeq  \frac{4GJ}{c^2 r} \sin^2\theta
\label{eq:gtphi}
\end{equation}
In general relativity, particles follow a geodesic path that can be calculated from the metric and its derivatives. The $g_{t\phi}$ term will add a perturbation to trajectories that orbit a spinning mass. For the Earth as a source, the effect is small and is currently at the limits of experimental tests \cite{iorio}. However, this paper is not concerned with the path through space and time that a particle takes when it is in a rotating gravitational field, but on the interaction between the asymmetric potential and the internal attributes of a particle.

When considering possible sources for the Kerr metric a close object with a large total angular momentum is required; the obvious candidates are the Earth, Sun and Galaxy.  Table~\ref{tab:sources} shows the relative magnitude of the $g_{t\phi}$ term with the Earth, Sun and Galaxy as the source. For the Galactic terms, a simple approximation based on the velocity of the sun and distance to the center of the galaxy is used\cite{feast_whitelock}; the exact values are uncertain, but the relative significance of the Earth, Solar and Galactic terms is clear enough.

The Kerr metric is well suited to the symmetry of a rotating mass and the $g_{t\phi}$ term clearly shows the asymmetry. However $g_{t\phi}$ is just one component of a tensor and its value will vary from one inertial frame to another and will depend upon the coordinates used - it can even be zero in some reference frames. By contrast the asymmetry in the curvature is a property of space time and it can be quantified independently of coordinates. The $\phi$ and $t$ coordinates used in the Kerr metric define two invariant vector fields, $\partial_\phi$ and $\partial_t$. These are called Killing vectors (see for example \cite{MTW}). From the invariant vector fields we can construct the scalar product - which is therefore an invariant scalar field that quantifies the asymmetry in the curved spacetime. Because it is a scalar it is independent of coordinates and is the same in all inertial frames. By construction the scalar field is given by the right hand side of equation~\ref{eq:gtphi} it defines an invariant scalar field, which quantifies the P and T asymmetry of the gravitational potential:
\begin{equation}
\psi_{CP} =  \frac{4GJ}{c^2 r} \sin^2\theta
\label{eq:psiCP}
\end{equation}

It is hypothesised that the scalar field $\psi_{CP}$ of the galaxy, acting oppositely on particles and anti-particles, is responsible for the observed CP violation seen in terrestrial particle physics experiments.

The spatial variables, $r$ and $\theta$, have the same meaning as in spherical polar coordinate systems centred on the source. The hypothesis makes clear predictions for the radial and angular dependence of CP violation in relation to the galaxy, sun and Earth; and the relative magnitude of each contribution. It also describes a model of CP violation that will vary throughout the Universe. The scalar character of the field gives a Lorentz invariant effect as seen experimentally.

The magnitude of the asymmetry can be calculated by expressing the metric as a perturbation of an orthonormal flat Lorentz metric, $\eta_{ab}$, which has the simple diagonal form $(-1,1,1,1)$ in natural units $c=1$ and with indices $a$, $b$ ranging over the coordinates $ t, x, y, z$. Writing $g_{ab} = \eta_{ab} + h_{ab}$ The asymmetric perturbation is evident in $h_{tx} = 4 G J /c^3 r^2$. Which is also dominated by the galactic source with a magnitude of order $10^{-9}$ for the space time asymmetry. The effect of the term is to give a time dilation effect, but the nature of the effect is quite different to the well-known gravitational redshift. The latter is due to the $r$ dependence of the $g_{tt}$ term, which determines the rate at which a clock ticks in different places. It therefore requires a significant change in $r$ to be observable. By contrast the $g_{tx}$ term governs the observed duration of motion in the $x$-direction. For light-like motion the time dilation is calculated by setting $ds=0$ and solving for $dt$. The duration is changed by a factor of $1+h_{xt}$ regardless of the distance. If processes are taking place that have different left-right asymmetry in particles and antiparticles, then the asymmetric metric has the potential to add a time asymmetry $\bigcirc (10^{-9})$ to processes, including decay processes.

There is currently no model that can relate the magnitude of environmental asymmetry above to observed CP violation. Experimental CP violation ratios can vary from near unity (in the heavy mesons) to the limits of detection. Within the standard model, CP violation is accounted for with the mass mixing matrix and superimposed quantum states. The complex probability amplitudes mean that the probability of decay rates are not a sum of probabilities of decay routes, but can include oscillations and interference between decay channels and can be sensitive to small differences of energy or lifetime \cite{burcham_jobes}, such characteristics are intrinsically quantum mechanical and cannot be modeled by any known classical process. There is no single model independent number to quantify experimental CP violation - The most commonly quoted parameter independent of phase definitions is the Jarlskog, J parameter \cite{jarlskog85} with a value of $ 3 . 10^{-5}$ \cite{particledatagroup2010}.

It is not immediately obvious how the unambiguous distinctive predictions could be measured experimentally with current technology because terrestrial experiments will all have the same $r$ and $\theta$ co-ordinates. However, the character of the gravitational asymmetry is anisotropic - the direction of rotation being distinguished by the $\partial_\phi$ killing vector field. This suggests a second order effect where the magnitude of CP violation is greater for decays in the $\phi$ direction. Such an effect could be tested for a galactic source by using existing data, by plotting CP violation as a function of the orientation of the decay products with respect to the fixed stars. This would require spatial plots as a function of the sidereal time. The smaller solar term would have an orientation modulated by local time, while the much smaller terrestrial effect would have constant distribution in the laboratory frame, but with a latitude dependency.
\begin{table}
\caption{Comparison of the CP anisotropic term for Earth, Sun\cite{pijpers} and Galactic sources. The Galactic source has the dominant contribution.}
\label{tab:sources}
\begin{center}
\begin{tabular}{rrrr l}
  \hline
  & Earth&Sun&Galaxy&\\ [0.5ex]
   $r$ & $6.3.10^6$ &$1.5.10^{11}$ & $2.5.10^{20}$ & $m $\\
   $J$ & $7.10^{33}$&$ 2.10^{41}$ & $ 10^{66}$ & $kg\ m^2 s^{-1}$ \\
  $ g_{t \phi}$ & $ 3\ \ \ \ \ \ $&$3.10^{3\ }$ & $10^{20}$ & $ m^2 s^{-1}rad^{-1}$ \\
  $h_{tx}$ & $10^{-15} $& $10^{-14}$&$10^{-9}$&dimensionless \\ [1ex]
  \hline
\end{tabular}
\end{center}
\end{table}

\section{Conclusion}
Experimental tests or CP violation have an in-built asymmetry caused by the known gravitational potential of spinning masses. The Galaxy is the source with the largest effect. With a variety of motivations, it is predicted that the anisotropy in space time is responsible for the observed CP violation. The prediction has a clear signature for the magnitude of the effect with distance and latitude with respect to the source. For terrestrial experiments the effect could be manifested as an anisotropy in measured CP violation because the gravitational CP violating term distinguishes the direction of galactic rotation. The experiments have already been carried out, and evidence can be sought in the orientation, with respect to the fixed stars, of decay products in CP asymmetric events - or equivalently plotting CP violation as a function of decay product direction.

If the prediction is confirmed then Nature can be considered to be symmetric under time reversal and space inversion.

CP violation is seen as the key to explaining the matter asymmetry in the Universe, but the measured CP violation is inadequate to explain the Universe that we see today (see~\cite{harrison:1998} and references therein). The proposed mechanism for CP violation in this paper, has considerable implications for observed baryon asymmetry in the Universe. On the one hand it provides for a much stronger CP violation near massive spinning sources, but fails to give universal CP violation.


\end{document}